\documentclass{INTERSPEECH2023}
\usepackage{amsmath}
\usepackage{color, colortbl}
\usepackage{array,multirow,graphicx}
\usepackage{tikz}
\usetikzlibrary{decorations.pathreplacing}

\newcommand{\varA}[1]{{\operatorname{\mathrm{#1}}}}

\usepackage[belowskip=-8pt,aboveskip=5pt]{caption}
\usepackage{float}

\interspeechcameraready

\title{Two-Stage Voice Anonymization for Enhanced Privacy}
\name{Francesco Nespoli$^{1,3}$, Daniel Barreda$^1$, J{\"o}erg Bitzer$^2$, Patrick A. Naylor$^3$}
\address{
  $^1$Microsoft, London, UK\\
  $^2$Fraunhofer Institute for Digital Media Technology IDMT, Oldenburg, Germany\\
  $^3$Imperial College, London, UK}
\email{fnespoli@microsoft.com, daniel.almendrobarreda@microsoft.com, joerg.bitzer@idmt.fraunhofer.de, p.naylor@imperial.ac.uk}

\let\OLDthebibliography\thebibliography
\renewcommand\thebibliography[1]{
  \OLDthebibliography{#1}
  \setlength{\parskip}{0pt}
  \setlength{\itemsep}{1.8pt plus 1ex}
}
\begin{document}

\maketitle
 
\begin{abstract}
  In recent years, the need for privacy preservation when manipulating or storing personal data, including speech , has become a major issue. In this paper, we present a system addressing the speaker-level anonymization problem. We propose and evaluate a two-stage anonymization pipeline exploiting a state-of-the-art anonymization model described in the Voice Privacy Challenge 2022 in combination with a zero-shot voice conversion architecture able to capture speaker characteristics from a few seconds of speech. We show this architecture can lead to strong privacy preservation while preserving pitch information. Finally, we propose a new compressed metric to evaluate anonymization systems in privacy scenarios with different constraints on privacy and utility.   
\end{abstract}


\section{Introduction}

    Speech is a primary modality for humans to communicate. In recent years, speech technologies enabled effective human-machine interaction making it possible to control systems and devices with speech \cite{humanmachine}. If, on one side, voice assistants and smart speakers facilitate daily tasks \cite{BetterLife}, then, on the other, these technologies raise privacy concerns for the public and policy makers \cite{privacyconcerns}. Such concerns come from the fact that speech contains significant personal identifiable information (PII) both in the semantic and acoustic domain. Specifically, personal identifiers such as full name, social security number or geographical positioning can alone allow speaker identification. Moreover, voice characteristics such as prosody, speaking rate, accent and intonation inherently contain a variety of PII such as personality, physical characteristics, emotional state, age and gender that can be identified \cite{Kröger2020} and therefore used for malicious privacy attacks. In this context, suppressing PII in speech signals would improve privacy. Considering a situation in which personal identifiers are not present or has been obfuscated, initial acoustic privacy protection approaches explored several research directions such as extracting privacy-preserving features \cite{six}, working with encrypted speech signals \cite{seven}, learning adversarial features \cite{nine}, or performing federated learning \cite{federated}.  However, feature or model-level privacy protection have limitations. Privacy preserving features can, in principle, be used for any downstream application but there is no guarantee they retain any useful information to efficiently address the new task. Encryption, although allowing data manipulation in the encrypted domain, introduces significant computational overheads. Adversarial features have been reported to increase privacy in closed-set classification scenarios but lack in generalization \cite{nine}. Federated models, despite avoiding direct access to the data, can leak original data information through the higher level representation used for learning (e.g., local gradients) \cite{federLeak}. Therefore, state-of-the-art anonymization systems are based on the idea of disentangling the speaker identity information from the linguistic and prosodic content thus producing synthetic utterances in which the speaker identity has been altered \cite{voicepriv2020},~\cite{xvecdesignchoice},~\cite{diffprivspeakanon} while other speech characteristics, possibly important for downstream tasks, are preserved. Although this general approach has been shown to be effective \cite{voicepriv2020}, the potential for concrete attacks is quite large. Specifically, speaker information can leak into linguistic and prosodic features \cite{leak}, propagate to the modified speech and be used by an attacker to identify the speaker. 
    
    One effective technique for speech anonymization employs automatic speech recognition (ASR) to transcribe speech followed by a text-to-speech (TTS) system that re-synthesizes audio signals from text transcriptions. This ASR+TTS method protects the vocal characteristics of the speaker's identity but completely destroys the original prosodic attributes such as intonation, stress and rhythm \cite{sarina}. Moreover, the incorrect linguistic content induced by any ASR error and the very limited variability of TTS-synthesized speech outputs, mainly due to few available voices, lead to poor results when applying ASR+TTS on downstream tasks \cite{poor1},~\cite{poor2}.  Another direction recently investigated by \cite{MultiStage} and \cite{2stgehighersecurity}, demonstrated that a cascade of signal-based anonymization modules results in higher anonymization scores compared to single-stage processes specifically in the case of decryption attacks \cite{2stgehighersecurity}. Based on these findings, we investigate the anonymization capabilities of a fully-neural pipeline based on voice conversion ($\varA{VC}$) targeting higher privacy enforcement in the context of voice privacy protection. The system combines, in a two-step procedure, a zero-shot voice conversion ($\varA{ZS-VC}$) block and the baseline system of the Voice Privacy Challenge \cite{voicepriv2020}. Privacy and utility scores have been measured as the equal error rate (EER) of an automatic speaker verification (ASV) system and the word error rate (WER) of an ASR model respectively. Furthermore, we computed a minimal secondary evaluation metric, specifically a lower-bound for the pitch correlation between original and synthesised utterances. Both the anonymization and scoring pipelines follow the framework of \cite{voice2022} in terms of datasets allowed for training and testing the anonymization models.  
    
    

\section{Proposed Model}

     In \cite{2stgehighersecurity}, the authors show a combination of multiple speech modifications enhances privacy. Based on those results, we propose a cascade of two deep learning-based voice conversion systems which target specific situations with different anonymazation requirements. Figure~1 provides a schematic overview of the components we used in the two-stage system. 
    \begin{figure}[ht]
    \centerline{\includegraphics[scale=0.255]{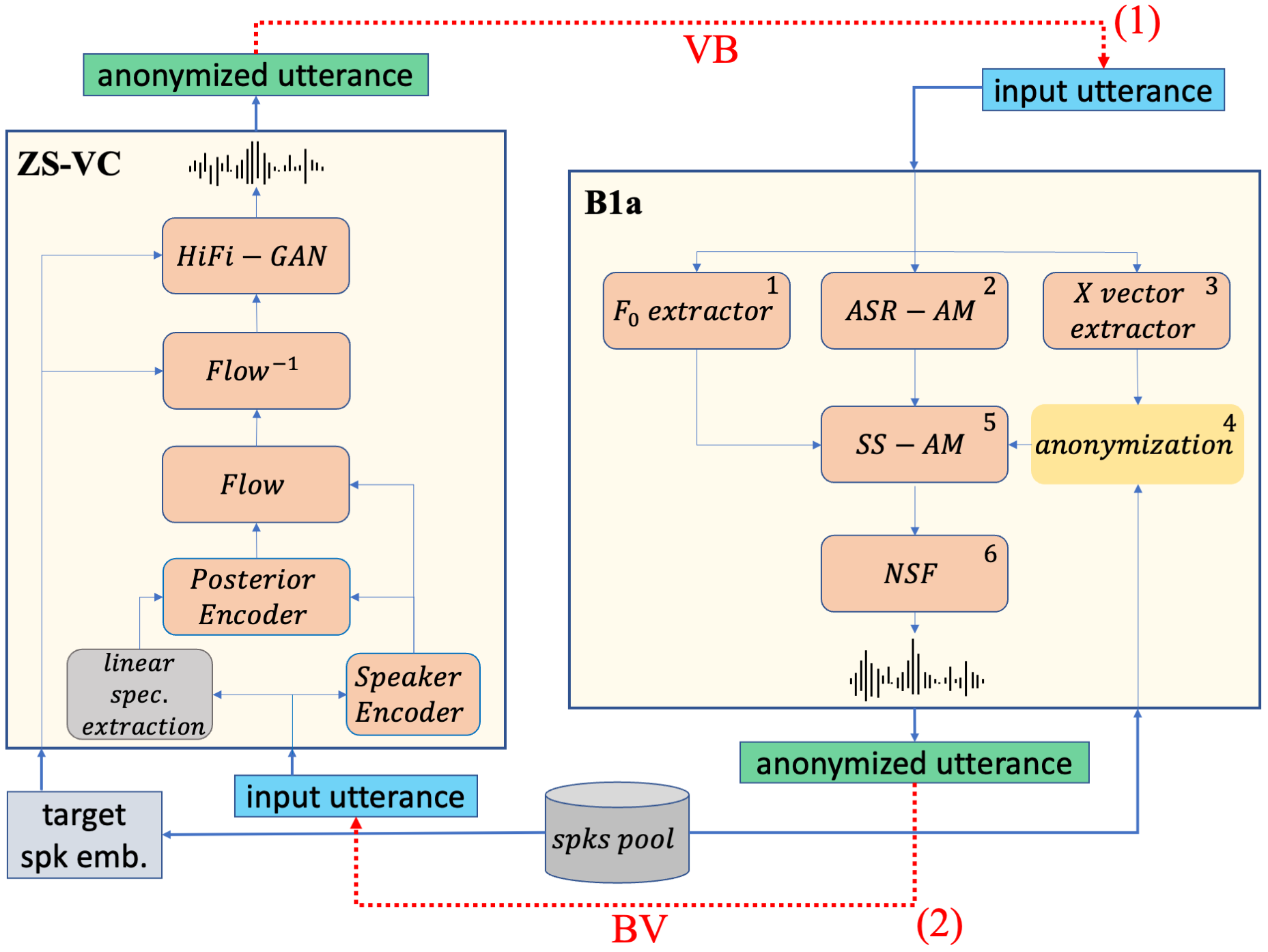}}
     \caption{Overview of voice anonymization components: zero-shot voice conversion ($\varA{ZS-VC}$) and baseline system ($\mathrm{B1a}$). The red dotted lines $\mathrm{(1)}$ and $\mathrm{(2)}$, are mutually exclusive and refer, in the Experiment section, to system $\mathit{VB}$ and $BV$ respectively.}
    \label{fig1}
    \end{figure}The first component $\varA{ZS-VC}$ is a zero-shot any-to-any voice conversion system \cite{yourtts} whereas the second is the baseline $\mathrm{B1a}$ from \cite{voicepriv2020}. Speech anonymization is achieved by altering the original speaker identity. To do this, both components rely on a set of external speakers (\textit{spks pool}, Fig.~1) utterances. Speakers in \textit{spks pool} are not included in the set to be anonymized: they are used as targets to alter the original speaker identity. Target speaker selection is handled differently by the two components and this is described in detail in the next sections. The red dotted lines in Fig~1 are mutually exclusive: when (1) is active, $\varA{ZS-VC}$ is applied first; when (2) is operating, $\mathrm{B1a}$ is the first stage of the two-stage anonymization process.

\subsection{Zero-Shot Voice Conversion}

    This stage exploits YourTTS \cite{yourtts} a multi-speaker TTS architecture with multilingual capabilities based on variational inference and adversarial learning \cite{vits}. The model has five main components: text encoder, posterior encoder, alignment stage, an invertible flow-based decoder and a vocoder \cite{hifi}. At training time, the posterior encoder receives as input the linear spectrogram and outputs a latent representation $Z$. This is used by the vocoder and by the flow-based decoder whose output $Z\textsubscript{p}$ is then aligned with the text encoder representation with monotonic alignment search (MAS) \cite{glowtts}. At inference time, the text encoder generates the alignment $Z\textsubscript{p}$ which is then used by the inverted flow decoder to produce $Z$ which, when input into the vocoder, generates the audio. Due to the fact that all five main components are conditioned on the speaker embedding, this same architecture can be used for voice conversion. In this case, given that at inference time we have an utterance instead of text, $Z\textsubscript{p}$ is obtained from the output of the chain posterior encoder and direct flow, which are conditioned on the original speaker embedding. At this point, by conditioning the inverted flow and the vocoder on the target speaker embedding, we can achieve effective $\varA{VC}$. Here, the speaker encoder was implemented with the H/ASP architecture \cite{HASP} based on ResNet-34. We explored two modalities for target speaker selection addressing different privacy-to-utility ratios. We hypothesize random target selection with gender preservation might allow $\varA{ZS-VC}$ to generate higher quality speech as observed in other voice conversion approaches \cite{autovc}. This could lead to fewer ASR errors and therefore lower WER. Contrarily, we argue that relaxing the gender constrain might result in higher EERs due to the greater mismatch between target and original utterance which might heavily affect the ASV embedding model.

\subsection{X-vector Speaker Anonymization}

    This step involves state-of-the-art techniques for extracting speaker identity, linguistic and prosodic information from the original speech signal and re-synthesizing a new utterance after modifying the speaker content. Blocks~1-6 in Fig.~1 come from \cite{voicepriv2020}. They consist of a first stage for speaker and acoustics information extraction (blocks~1-3) and a second re-synthesis stage (blocks~4-6), with the idea of combining original speaker acoustics with a new speaker embedding for efficient voice anonymization. The first stage comprises a fundamental frequency ($\mathrm{F_0}$) extractor (block~1) based on \cite{pitchpraat} a time-delay neural network (TDNN) ASR to extract bottle neck (BN) features modeling speech acoustics (block~2) and a speaker encoder (block~3) computing x-vectors \cite{xvector} with a second factorized-TDNN architecture. Anonymized speech is synthesised in two steps. First, a speech synthesis acoustic model (SS AM - Block~5) computes  Mel-filterbank features given $\mathrm{F_0}$, BN features and the anonymized x-vector. Second, a neural source-filter (NSF - block~6) generates a waveform from $\mathrm{F_0}$, anonymized x-vector and Mel-filterbank. In this case, the anonymized x-vector has been obtained using the following distance-based rule. First, a set of \textit{N} closest speaker embeddings is selected for each original speaker representation. Then, a subset \textit{M} is randomly sampled from \textit{N} and averaged to produce the target speaker embedding (\textit{N}=200, \textit{M}=20).
    
\section{Experiments}

    \begin{table*}[t]
    \centering
    \caption{Primary anonymization metric ($\mathrm{EER}$, higher is better) results for development (light gray) and test (dark gray) sets. $\mathrm{Orig}$: Non-anonymized original data. $\mathrm{B1a}$: baseline from~\cite{voicepriv2020}. $V_R$: $\varA{ZS-VC}$ with random target speaker selection. $V_{GP}$: $\varA{ZS-VC}$ with random target speaker selection and gender preservation. $\mathit{BV}$: $\mathrm{B1a}$ followed by $\varA{ZS-VC}$. $\mathit{VB}$: $\varA{ZS-VC}$ followed by $\mathrm{B1a}$. $\mathrm{VCTK}$ subsets comprise a set of utterances with same ($\mathrm{comm.}$) and different ($\mathrm{diff.}$) linguistic content across the speakers.}

    \definecolor{Gray}{gray}{0.9}
    \definecolor{Gray_2}{gray}{0.75}
    
    \label{tab:Table 3}
    \begin{tabular}{ccccccccc}
    \toprule
    \textbf{Set}&\textbf{Gender}&\textbf{Weight}&\multicolumn{6}{c}{\textbf{EER[\%]}} \tikz [remember picture] \node (rightmark) {}; \\
    \midrule
    &&&Orig&$\mathrm{B1a}$&$V_{R}$&$V_{GP}$&$\mathit{BV}$&$\mathit{VB}$ \\ 
    \cline{1-9}
    \rowcolor{Gray}
    \rowcolor{Gray}
    Libri-dev    & female  & 0.25 & 4.12 & 14.6 & 21.6 & 22.3 & 32.95  & 52.6  \\
    \rowcolor{Gray}
                 & male    & 0.25 & 0.93 & 10.2 & 16.2 & 18.3 & 34.6   & 43.9                                   \\
    \rowcolor{Gray}
    Vctk-dev     & female  & 0.20 & 0.84 & 9.1 & 26.9 & 30.4 & 35.1   & 53.3                                    \\
    \rowcolor{Gray}
    
    diff.        & male    & 0.20 & 0.64 & 8.1 & 27.9 & 30.0 & 19.52  & 41.4                                       \\
    \rowcolor{Gray}
    Vctk-dev     & female  & 0.05 & 0.87 & 10.2 & 26.8 & 28.7 & 25.6   & 42.4                                    \\
    \rowcolor{Gray}
        comm.    & male    & 0.05 & 0.58 & 9.7 & 30.5 & 29.1 & 32.48  & 51.3                                     \\
    
    \midrule
    \boldsymbol{$Avg^W dev$} & &  & 1.63 & 10.64 & 23.3 & 25.1 & 30.7 & 47.8\\
    \midrule
    
    \rowcolor{Gray_2}
    Libri-test   & female  & 0.25 & 2.55 & 12.7 & 18.8   & 14.8  & 31.8 & 46.0   \\
    \rowcolor{Gray_2}
                 & male    & 0.25 & 0.43 & 10.5 & 11.4   & 11.8  & 30.0 & 48.5                                      \\
    \rowcolor{Gray_2}
    Vctk-test    & female  & 0.20 & 1.59 & 14.7 &  30.4  & 32.3  & 29.8 & 35.2                                   \\
    \rowcolor{Gray_2}
       diff.     & male    & 0.20 & 0.97 & 12.2 &  27.1  & 28.3  & 32.9 & 39.0                                     \\
    \rowcolor{Gray_2}
    Vctk-test    & female  & 0.05 & 0.34 & 13.8 &  29.5  & 30.9  & 27.8 & 41.6                                       \\
    \rowcolor{Gray_2}
       comm.     & male    & 0.05 & 0.28 & 7.1  &  27.4  & 29.4  & 31.3 & 48.9                                                          \\
    
    \midrule
    \boldsymbol{$Avg^W test$} & &  & 1.29 & 12.2 & 21.9 & 21.8 & 31.0 & 43.0\\
    \bottomrule
    \end{tabular}
    \end{table*}

\subsection{Data}
    In the experiments we used the same training and testing subsets listed in \cite{voice2022} with the same constraints. Specifically, $\varA{ZS-VC}$ was trained first on \textit{LibriTTS-100-clean} \cite{zen19_interspeech} for $7\times10^5$  steps with batch size 52 and then fine-tuned on \textit{LibriTTS-500-others} for further $3\times10^5$ epochs. Both \textit{LibriTTS} subsets were resampled at 16kHz and RMS normalized with target level -27dB \cite{yourtts}. Furthermore, \cite{silero} was used to remove silences. The speakers pool in Fig.~1 coincides with the \textit{LibriTTS-train-other-500} subset and it was used to extract target speaker representations. First, a speaker embedding was computed for each utterance in the pool. Second, the target speaker representation was calculated by averaging all the speaker embeddings for each speaker. This process was computed for both H/ASP in $\varA{ZS-VC}$ and the TDNN x-vector extractor in $\mathrm{B1a}$. Differently from \cite{voicepriv2020}, we employed pre-training for both the ASV and ASR models. Specifically, instead of training the models from scratch only on the anonymized \textit{LibriSpeech-360-clean} subset~\cite{LibriSpeech} ($LS_{anon}^{360}$), we first pre-trained the ASV speaker embedding model on VoxCeleb~1,2~\cite{Chung18b} subsets following the recommendations of~\cite{speechbrain} and then fine-tuned on $LS_{anon}^{360}$ for 20 more epochs. The ASR model was pre-trained on \textit{LibriTTS-100-clean} and \textit{LibriTTS-500-others} for 60 epochs and fine-tuned $LS_{anon}^{360}$ for 30 more epochs. We used Adam optimizer with initial learning rate (\textit{lr}) 0.001, \textit{lr}-scheduler from \cite{scheduler} and batch size of 128. Pre-trainig and fine tuning on $LS_{anon}^{360}$ was conducted separately for each anonymization condition for both ASR and ASV. Finally, the anonymizaton results were tested on \textit{clean} development and test sets of \textit{LibriSpeech} and on a subset of \textrm{VCTK} \cite{vctk} obtained following the same procedures as in~\cite{voice2022}. 

\subsection{Metrics}
    The two primary scores to evaluate an anonimization system are the EER and the WER \cite{voice2022}. The first assesses the privacy protection capability of the anonymization pipeline whereas the second measures the utility of the anonymized speech to perform downstream tasks. Given a generic biometric authentication system $\mathrm{G}$, $R_\mathrm{fa}^\mathrm{G}(\theta)$ and $R_\mathrm{fr}^\mathrm{G}(\theta)$ are the false acceptance and false rejection rates at a given decision threshold $\theta$. The EER corresponds to the rate at which $R_\mathrm{fa}^\mathrm{G}(\theta)=R_\mathrm{fr}^\mathrm{G}(\theta)$. The WER is calculated from the ASR output transcription as \[ \mathrm{WER} = \frac{N_{\mathrm{sub}}+N_{\mathrm{ins}}+N_{\mathrm{del}}}{N_{\mathrm{tok}}} \] where $N_{\mathrm{sub}},N_{\mathrm{ins}},N_{\mathrm{del}}$ are the number of substitutions, insertions and deletions in the ASR output and $N_{\mathrm{tok}}$ is the number of tokens in  the reference transcript. Moreover, it has been shown that when manipulating speech data to enhance privacy (higher EER), the utility of the anonymized signals drops (higher WER)~\cite{voicepriv2020}. Therefore, the choice of the most appropriate anonymization system heavily depends on the specific application: when privacy is paramount, large WER degradation might be tolerated in favour of privacy. However, for other applications when even small WER variations heavily impact the performance, an anonymization system with a limited WER deterioration might be the best choice despite its lower privacy protection capability. ~Here, we propose the privacy-to-utility trade off ($\mathrm{PU_{tr}}$), a compressed metric combining the primary anonymization metric and primary utility score, designed to evaluate the anonymization system at different operating points. Given $\mathrm{WER_i}, \mathrm{EER_i} \in (0,1], i\in\{0,1\}$ denotes the metrics calculated on original ($i=0$) or anonymized ($i=1$) utterances. We define 
    \vspace*{-0.5\baselineskip}
    \[\mathrm{PU_{tr}}=\lambda\frac{\log\left(1+\frac{\mathrm{WER_1}}{\mathrm{WER_0}}\right)}{\log\left(1+\frac{1}{\mathrm{WER_0}}\right)}-(1-\lambda)\frac{\log\left(1+\frac{\mathrm{EER_1}}{\mathrm{EER_0}}\right)}{\log\left(1+\frac{1}{\mathrm{EER_0}}\right)}\] where $\lambda\in[0,1]$ controls the trade off between WER and EER. $\mathrm{PU_{tr}}\in[-1,1]$ and a lower value indicates a more favorable trade-off at a specific operational point $\lambda$. Furthermore, we calculated $\rho^{F_0}$, also called pitch correlation metric, to assesses intonation preservation of the anonymization process. First, $\mathrm{F_0}$ was extracted from each utterance using \cite{pitchpraat}. Then, $\rho^{\mathrm{F_0}}$ was computed as the Pearson correlation between the $\mathrm{F_0}$ of original and anonymized speech.


\subsection{Scoring systems}
    In our experiments, the attacker is defined as \textit{Semi-Informed} \cite{voicepriv2020}. This means it has full knowledge of the anonymization system, but can not access the mapping between original and anonymized speakers. In this condition, the attacker anonymizes the training set \textit{LibriSpeech-360-clean} by selecting a different target speaker for each utterance in the dataset and fine-tunes the ASV model on these data. The speaker verification model we used is an x-vector TDNN-based speaker encoder coupled with a probabilistic linear discriminant analysis (PLDA) classifier \cite{plda}. The system uses the same verification files in Kaldi format as \cite{voice2022} in which enrollment and trial utterances have been anonymized with different target speakers. We employed an ASR system based on a transformer acoustic model encoder and a joint transformer decoder with connectionist temporal classification (CTC) \cite{CTC}, with decoding stage integrating also CTC probabilities. The ASR and ASV models were implemented with \cite{speechbrain}. 

\section{Results and Discussion}


\subsection{Primary Anonymization Metric}
    Privacy protection capabilities of each model have been evaluated with the EER metric. Results for each system have been reported in Table~1. Here, ZS-VC on its own provides higher protection compared to the baseline system scoring a 2-fold increment in the EER. Moreover, the concatenation of the two anonymization systems greatly enhances privacy protection of stand-alone pipelines with $\mathit{VB}$ scoring close to perfect anonymization (EER=50\%) for many test and development sets. Finally, preserving the gender information in the original-to-target mapping for ZS-VC appears to lead to greater anonymization results ($V_{GP}$ column, Table~1) when compared to random target selection ($V_{R}$ column, Table~1).

\subsection{Primary Utility Metric}

    We assessed linguistic information preservation by calculating the WER with the ASR model fine-tuned separately for each anonymization condition. Table~2 summarises the scores for each anonymization stage. As expected, all anonymization processes degrade ASR performance. However, in the case of ZS-VC, preserving the gender of the original speakers after the conversion, produces a 14.7\% WER reduction (on average between development and test sets) when compared to random gender mapping. Moreover, two-stage processing ($\mathit{BV}$ and $\mathit{VB}$ in Table~2), although degrading WER with respect to single stage processes, can achieve better utility scores by employing ZS-VC as the second stage of the anonymization pipeline ($\mathit{BV}$).  

    \begin{table}[ht]
    \vspace*{+3pt}
    \centering
    \caption{Primary utility metric (WER, lower is better).}
    \label{tab:Table 4}
    \begin{tabular}{ccccccc}
    \toprule
    \textbf{Set}&\multicolumn{6}{c}{\textbf{WER[\%]}} \\
    \midrule
    &Orig.&$B_{1a}$&$V_{R}$&$V_{GP}$&$\mathit{BV}$&$\mathit{VB}$ \\
    \cline{2-7}
    \cr
    
    Libri-dev   & 2.33  &  2.77 &  4.22 &  4.04 & 4.72  & 6.51     \\
    Vctk-dev    & 8.21  & 9.59  &  14.58 &  13.95 & 16.24 & 19.39              \\
    
    \midrule
    \boldsymbol{$Avg~dev$}  & 5.27 & 6.18 &  9.4 & 9.0 & 10.5 & 12.95 \\
    \midrule
    
    Libri-test  & 2.47  &  2.85 &  3.84 &  3.78 & 4.17  & 7.19    \\
    Vctk-test   & 7.63  & 9.39  &  13.65  &   9.86 & 9.63 & 18.12            \\
                                                     \\
    \midrule
    \boldsymbol{$Avg~test$}    & 5.1 & 6.12 & 8.75 & 6.82 & 6.92 & 12.7\\                                                
    \bottomrule
    \end{tabular}
    \end{table}
    
\subsection{Privacy Utility Trade-Off}
    One of the outstanding problems of privacy evaluation is that two interplaying quantities need to be evaluated at the same time. Specifically, when privacy improves (higher EER), it comes at the cost of WER degradation (Table~1,2). Here, we suggest to combine these two measures into a compressed metric $\mathrm{PU_{tr}}$.
            \vspace*{-\baselineskip}
            \begin{figure}[ht]
            \hspace{-1.2em}
            \centerline{\includegraphics[scale=0.235]{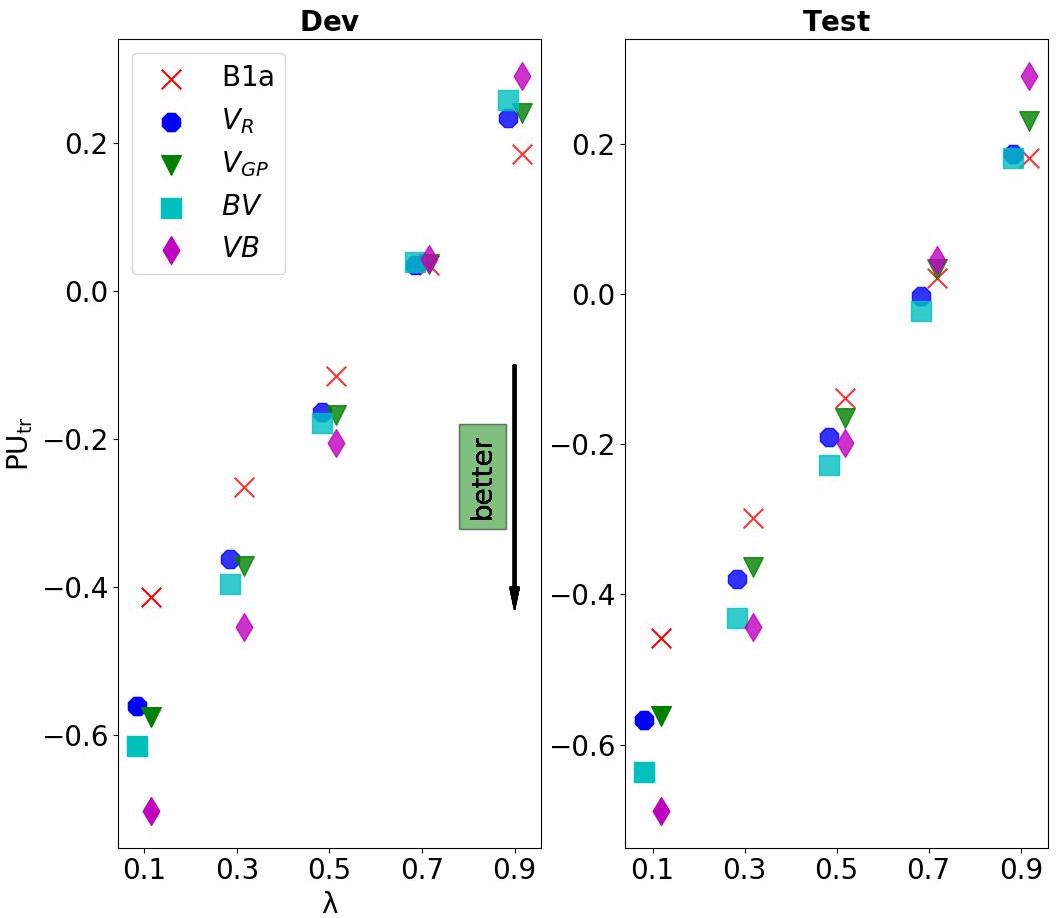}}
             \caption{Privacy-to-utility trade-off for development ($\mathbf{Dev}$) and test ($\mathbf{Test}$) sets evaluated for $\lambda\in[0.1, 0.3, 0.5, 0.7, 0.9]$.}
            \label{fig2}
            \end{figure}
    Figure~2 shows $\mathrm{PU_{tr}}$ results for each anomymization system and for different values of $\lambda$. In privacy applications where a low WER is crucial ($\lambda=0.9$) $\mathrm{B1a}$ results to be the best choice. However, for applications in which users require stronger privacy and can tolerate higher WER increments (lower values of $\lambda$) $\varA{ZS-VC}$ and two-stage processes become a better choice in terms of $\mathrm{PU_{tr}}$.

\subsection{Intonation Preservation}

    We tested intonation preservation by calculating $\rho^{\mathrm{F_0}}$ from original and anonymized utterances. Although $\mathrm{F_0}$ has been proven to incorporate speaker information \cite{pitchspeaker}, all ZS-VC models display high $\rho^{\mathrm{F_0}}$ while improving the EER, showing that it is possible to achieve strong anonymization while maintaining prosody information (Table~3). Specifically, two-stage methods can achieve EERs comparable with state-of-the-art ASR+TTS anonymization systems \cite{sarina} while greatly improving $\rho^{\mathrm{F_0}}$. This is particularly important when ASR models trained on anonymized data need to be applied to real speech \cite{poor1},~\cite{poor2}.

    \begin{table}[ht]
    \vspace*{+3pt}

    \centering
    \caption{Pitch correlation ($\rho^{\mathrm{F_0}}$, higher is better). Average results computed with the same weights as in Table~1.}
    \label{tab:Table 5}
    \begin{tabular}{ccccccc}
    \toprule
    \textbf{Set}&\textbf{Gnd}&\multicolumn{5}{c}{\textbf{$\rho^{\mathrm{F_0}}$}} \\
    \midrule
    &&$B_{1a}$&$V_{R}$&$V_{GP}$&$\mathit{BV}$&$\mathit{VB}$ \\ 
    \cline{3-7}
    \cr
    
    Libri-dev      & F        & 0.77    &  0.8    &  0.81  & 0.8   & 0.75  \\
                   & M        & 0.73    &  0.78   &  0.75  & 0.77  & 0.75  \\
    Vctk-dev       & F        & 0.84    &  0.81   &  0.85  & 0.82  & 0.79  \\
      comm.        & M        & 0.78    &  0.78   &  0.77  & 0.78  & 0.74  \\
    Vctk-dev       & F        & 0.79    &  0.77   &  0.81  & 0.79  & 0.76  \\ 
       diff.       & M        & 0.72    &  0.75   &  0.74  & 0.75  & 0.71  \\
    \midrule
    \textbf{Avg dev}   &      & 0.78    &  0.79   &  0.79  & 0.79  & 0.76       \\ 
    \midrule
    
    Libri-test     & F        & 0.77    &  0.8    &  0.85  & 0.81  & 0.76  \\
                   & M        & 0.69    &  0.71   &  0.72  & 0.73  & 0.67  \\
    Vctk-test      & F        & 0.84    &  0.83   &  0.86  & 0.82  & 0.81  \\
        comm.      & M        & 0.79    &  0.78   &  0.78  & 0.78  & 0.76  \\
    Vctk-test      & F        & 0.79    &  0.8    &  0.83  & 0.79  & 0.77  \\
        diff.      & M        & 0.70    &  0.74   &  0.72  & 0.73  & 0.71  \\
    \midrule
    \textbf{Avg test} &       & 0.77    &  0.78   &  0.80 & 0.78  & 0.75 \\ 
    \bottomrule
    \end{tabular}
    \end{table}

\section{Conclusions}
In this paper we present a novel anonymization pipeline cascading two fully-neural anonymization systems. This was achieved using a combination of zero-shot voice conversion and a state-of-the-art anonymization model. Results show that two-stage processes can preserve prosodic information while concealing speaker identity with EER scores comparable with ASR+TTS methods.  We introduce $\mathrm{PU_{tr}}$, a compressed metric to evaluate the anonymization models for privacy applications with different WER and EER constrains and we showed that with this new score, the choice of the best anonymization technique can be tuned with the $\lambda$ parameter according to specific anonymization requirements in terms of WER and EER.

\section{Acknowledgements}
This project was funded by the European Union’s Horizon 2020 program under the Marie Skłodowska-Curie grant No 956369.


\bibliographystyle{IEEEtran}
\bibliography{sapstrings,new_bib}

\end{document}